# DATI COVID E LEGGE DI BENFORD

*di Filippo Elba[1]*

**Abstract**

È qui proposta un'applicazione della legge di Benford (o della prima cifra) ai dati giornalieri per Paese dei nuovi casi registrati di contagi da Covid 19. La fonte dati utilizzati è https://ourworldindata.org/coronavirus-source-data. L'intento principale di questo scritto è quello di verificare se la distribuzione delle prime cifre dei numeri relativi ai casi giornalieri registrati rispetta la distribuzione della legge di Benford. Seguono alcune brevi osservazioni sui risultati ottenuti.

**Parole chiave:** legge di Benford; Covid 19.

**Legge di Benford: caratteristiche**

La legge di Benford (o legge della prima cifra) è una distribuzione di probabilità che descrive la probabilità che un numero, presente in una raccolta di dati reali (popolazione dei comuni, quotazione delle azioni, costanti fisiche o matematiche, etc.), cominci con una data cifra. Tale funzione di probabilità è uguale a:

$$P(d) = \log_{10}(d+1) - \log_{10}(d) = \log_{10}\left(1 + \frac{1}{d}\right) \quad d = 1, 2, \ldots, 9 \qquad (1)[2]$$

La particolarità di quanto affermato da questa legge sta nel fatto che, stanti alcune condizioni, le probabilità che un numero *x* avrebbe di iniziare con una cifra compresa tra 1 e 9 non sarebbero tra loro uguali (ossia dello 0,11 ca.).

Negli anni '30, Benford[3] dimostra ciò ricorrendo a più di 20.000 casi concernenti le più disparate serie di dati. In ogni verifica da lui effettuata, le frequenze con cui i numeri degli insiemi presi in esame iniziano con la cifra "1" sono di gran lunga maggiori rispetto a tutte le altre e sempre vicine alla quota di un terzo rispetto al totale. Fenomeno che non sembra riscontrarsi quando l'analisi passa alle cifre successive. In questi casi, infatti, le frequenze relative sono più uniformi rispetto alla probabilità dello 0,10 (dalla seconda cifra in avanti c'è da aggiungere la possibilità che essa possa assumere valore "0").

Per le cifre successive alla prima varrebbe la seguente funzione di probabilità:

$$P(d) = \sum_{k=10^{n-2}}^{10^{n-1}-1} \log_{10}\left(1 + \frac{1}{10k+d}\right) \quad con\ n > 1;\ d = 0, 1, \ldots, 9 \qquad (2)\ [4]$$

Tale legge sarebbe indifferente rispetto alla scala di misurazione adottata. Dimostrazioni effettuate da Benford e da altri confermano questa proprietà[5].

---

[1] Assistente alla ricerca, Dipartimento di Scienze per l'Economia e l'Impresa, Università degli Studi di Firenze.

[2] La (1) è applicabile anche a stringhe di cifre. In questo caso *d* assume il valore della stringa d'interesse.

[3] In realtà, già prima di Benford, pare che questa legge fosse stata scoperta, dalla fine del XIX secolo, dal matematico e astronomo Simon Newcomb. Egli notò che, le pagine dei libri con le tabelle dei logaritmi riferite a numeri aventi "1" come prima cifra, erano più sporche delle altre. Ne dedusse che, probabilmente, ciò dipendesse dal fatto che venivano usate più spesso. Venne controargomentato, tuttavia, che in qualsiasi libro, al quale si accede alle pagine in modo sequenziale, le prime tra esse sarebbero state sempre più usate delle ultime.

[4] Con *n* che rappresenta la posizione occupata dalla cifra *d* nella stringa numerica (sui valori assunti delle frequenze relative riferite alla seconda cifra in base alla (2): Tab.2).

[5] MathWord – Benford's Law (http://mathworld.wolfram.com/BenfordsLaw.html).

Purché le unità di misura adottate non siano espresse in base al sistema binario, la maggior frequenza con cui si registrano numeri con la prima cifra piccola sarebbe sempre empiricamente confermata.

Affinché la frequenza delle prime cifre di dati riferiti a un qualsiasi insieme siano quanto più vicini possibili alle probabilità benfordiane, è determinante il rispetto di alcuni "paletti":

- l'insieme dei numeri deve essere scelto sulla base di una variabile casuale;
- tutte le cifre da 1 a 9 devono avere stessa probabilità di poter essere al primo posto nei numeri che costituiscono l'insieme, senza alcun limite, anche inconsapevole;
- i numeri dell'insieme devono essere molti, evitando ripetizioni;
- l'insieme non deve essere costituito da numeri "assegnati" (per esempio, i numeri telefonici);
- il valore della media dei numeri dell'insieme deve essere maggiore rispetto alla mediana;
- i numeri dell'insieme considerato devono essere costituiti da più cifre (meglio se almeno quattro[6]).

È in particolare Hill (1995)[7] che si interessa alla determinazione delle caratteristiche che deve avere un insieme di numeri per poter meglio soddisfare le probabilità di distribuzione benfordiane: se si scelgono in modo casuale delle distribuzioni che godono della proprietà di essere invarianti rispetto alla scala di misurazione e dei campioni casuali siano presi da ciascuno di essi, le frequenze relative delle prime cifre dei numeri così ottenuti seguiranno, molto probabilmente, la legge di Benford. Sarebbero le distribuzioni di seconda generazione, quelle risultato della combinazione di altre distribuzioni, ad adattarsi meglio alle previsioni benfordiane.

Come già precedentemente sottolineato, quella di Benford è una legge empirica e, in quanto tale, difficile è cercarne un fondamento teorico. In *The Law of Anomalous Numbers* (1937), tuttavia, Benford rileva come molti fenomeni in natura si caratterizzerebbero per il fatto di seguire scale logaritmiche o geometriche piuttosto che aritmetiche. Tra gli esempi che egli cita: la crescita della sensibilità della retina alla brillantezza all'aumentare progressivo dell'illuminazione, dell'orecchio all'aumentare della rumorosità, la cognizione del tempo all'aumentare dell'età. In generale, negli ambiti più diversi fra loro (dalla medicina alla fisica atomica), tenderebbe sempre a presentarsi questa situazione, negando, perciò, la "naturalezza" del concetto di proporzionalità costante. Come lo stesso Bendford afferma alla fine del suo scritto "*small things are more numerous than large things, and there is a tendency for the step between sizes to be equal to a fixed fraction of the last preceding phenomenon or event*"[8].

**Alcuni casi pratici di applicazione della legge**

Diversi sono gli ambiti in cui si è cercato di applicare la legge di Benford: da indagini riguardanti presunti brogli elettorali, a controlli effettuati sui mercati finanziari durante la crisi del 2007-08[9] o relativi alla veridicità dei risultati pubblicati nelle riviste scientifiche[10].

Tra le applicazioni più note della legge di Benford ci sono quelle di Varian e di Nigrini.

---

[6]Affinché sia rispettata al meglio la distribuzione benfordiana sulla prima cifra, ideale sarebbe considerare numeri dell'ordine delle migliaia. In questo modo la variabile "prima cifra del numero" si configurerebbe come continua e non come discreta (cosa che accade, in particolare, per i numeri costituiti da unica cifra). Per una miglior descrizione di quanto detto: Benford F. (1937), *The Law of Anomalous Numbers,* Proceedings of the American Philosophical Society 78 (4), pagg. 551–572.

[7]Hill T. (1995), *A statistical derivation of the Significant-Digit Law,* Statistical Science 10 (4), pagg. 354-363.

[8]Benford F. (1937), *op. cit.*, pag. 571.

[9]Hofmarker P. e Hornik K. (2010), *First significant digit and the credit derivative market during the financial crisis,* Research Report Series / Department of Statistics and Mathematics,101. Institute for Statistics and Mathematics, WU Vienna University of Economics and Business, Vienna.

Diekmann A. (2004), *Not the first digit! Using Benford's Law to detect fraudolet scientific data*, Journal of applied statistics (http://128.118.178.162/eps/othr/papers/0507/0507001.pdf).

Nel 1972, Varian suggerí la possibilità di utilizzare questa legge per individuare eventuali falsificazioni nelle raccolte di dati usate per supportare decisioni politiche, basandosi sul presupposto che chi vuole addomesticare i dati sarebbe portato ad usare numeri distribuiti in modo non naturale, in contrasto, dunque, con le previsioni benfordiane. Comparando la frequenza relativa delle prime cifre di un insieme di numeri con le frequenze di Benford, si potrebbero evidenziare risultati anomali. Alla stessa maniera si può usare questo confronto per cercare falsificazioni in raccolte di dati riguardanti, per esempio, costi e entrate.

Altra applicazione conosciuta è quella fatta da Nigrini (1996)[11]. Dopo aver provato l'efficacia della legge di Benford su casi reali di frode accertata riferiti al 1992, egli studia come utilizzarla in maniera sistematica per testare il contenuto delle dichiarazioni dei redditi. Un caso peculiare, spesso ricordato dallo studioso americano per confermare la validità dell'utilizzo di questa legge nell'ambito dei controlli fiscali, si verifica nel 1993. In quell'anno, lo stato dell'Arizona cita in giudizio W. J. Nelson, accusato di aver cercato di defraudare lo stato per circa due milioni di dollari[12]. Nelson, manager dell'ufficio del Tesoro dell'Arizona, respinge le accuse dimostrando che non vi è alcun tipo di prova che possa confermare ciò. In realtà, le ventitré operazioni sospette imputategli registrano, il più delle volte, un ammontare di poco inferiore ai cento mila dollari (soglia passata la quale sarebbero previsti controlli aggiuntivi da parte di organismi appositi). Ciò implica un aumento del numero di volte in cui il "7", l'"8" e il "9" rappresentano la cifra iniziale delle operazioni effettuate da questo funzionario (90% il loro totale, in netta controtendenza rispetto a quanto previsto dalla legge della prima cifra).

**Legge di Benford sui dati dei contagi[13]**

Il presente lavoro presenta un'applicazione della legge di Benford ai numeri dei contagi giornalieri registrati in un gruppo di Paesi (Brasile, Cina, Francia, Germania, Giappone, Italia, Regno Unito, Stai Uniti). La rilevazione della banca dati utilizzata parte dal 31 dicembre 2019. Il calcolo delle frequenze per ogni Paese tiene conto dei soli giorni in cui si è registrato un numero di casi maggiore di zero. È per questo che per alcuni Paesi, vedi Cina, si dispone di quasi duecento osservazioni. Per altri, più recentemente toccati dal contagio, di poco più di centoquaranta (per esempio, Brasile e Francia).

Prima di confrontare le distribuzioni delle "prime cifre", è utile una breve panoramica sull'andamento dei casi giornalieri registrati nei Paesi che sono oggetto della presente analisi (Figura 1).

La Cina, primo Paese in assoluto a registrate casi di Covid 19, raggiunge un picco massimo a metà febbraio (i primi contagi si registrano circa un mese prima). Dopo una lunga decrescita che porta a quasi azzerare i casi tra metà maggio e i primi di giugno, affronta un nuovo incremento nel corso dell'estate, con numeri comunque molto al di sotto dei picchi invernali.

Brasile e Stati Uniti seguono un andamento molto simile tra loro: primi casi registrati tra fine febbraio e i primi di marzo, successivo andamento crescente che, seppur attenuato, sembra persistere anche nel luglio 2020.

I Paesi europei registrano i primi casi tra febbraio e marzo ma, dopo aver raggiunto i valori giornalieri più alti attorno alla prima metà di aprile, si avviano ad un lento ribasso, fatta eccezione per il Regno Unito in cui il rallentamento, seppur presente, sembra in qualche modo frenato.

L'altro Paese asiatico considerato, il Giappone, affronta i primi incrementi nella prima metà di marzo, raggiunge i suoi picchi, comunque limitati rispetto agli altri Paesi del gruppo, nella seconda metà di aprile e, dopo una flessione verso la fine di maggio, vede nuovamente incrementare il numero di nuovi casi nel corso dei mesi estivi.

---

[11]Nigrini M. (1996), *A Taxpayer Compliance Application of Benford's Law.,* J. Amer. Tax. Assoc. 18, 72-91, 1996.
[12]Nigrini M. (1999), *I've got your number*, Journal of Accountancy.
(http://www.journalofaccountancy.com/issues/1999/may/nigrini.htm).
[13] Il software utilizzato per le analisi è Microsoft Excel 2020.

Passando adesso al confronto delle distribuzioni delle prime cifre dei numeri dei casi giornalieri registrati nei Paesi del gruppo con la distribuzione di Benford (Figure 2:9) si nota, innanzitutto, come è assolutamente evidente che le distribuzioni siano molto differenti da quella uniforme. In tal senso è utile anche valutare gli esiti del test del chi-quadro in tabella 10, lato sinistro: in tutti i casi emerge come esista una differenza significativa tra le distribuzioni registrate e quella uniforme.

Ovunque è invece più evidente la maggior frequenza di numeri che iniziano con cifre basse. In tre casi (Germania, Italia e Stati Uniti) è maggiore la frequenza dei numeri che iniziano con il "2", sebbene, di fondo, l'andamento previsto dalla distribuzione di Benford resti rispettata.

Da una valutazione grafica generale, è la Cina il Paese che sembra adattarsi meglio alla distribuzione di Benford (per altro è quello con il numero di osservazioni giornaliere diverse da zero più elevato). Il Regno Unito è quello che sembra avere la distribuzione più "caotica", sebbene anche lì sia rispettato il principio di massima per cui le cifre più basse risultano più frequenti di quelle più elevate.

Il test del chi quadro (tabella 10, lato destro) conferma come, almeno per alcuni casi (quattro su otto), esista anche una evidenza statistica per poter affermare che esiste un qualche legame tra le distribuzioni osservate e quella di Benford.

1. **Nuovi casi giornalieri (scala log10)**

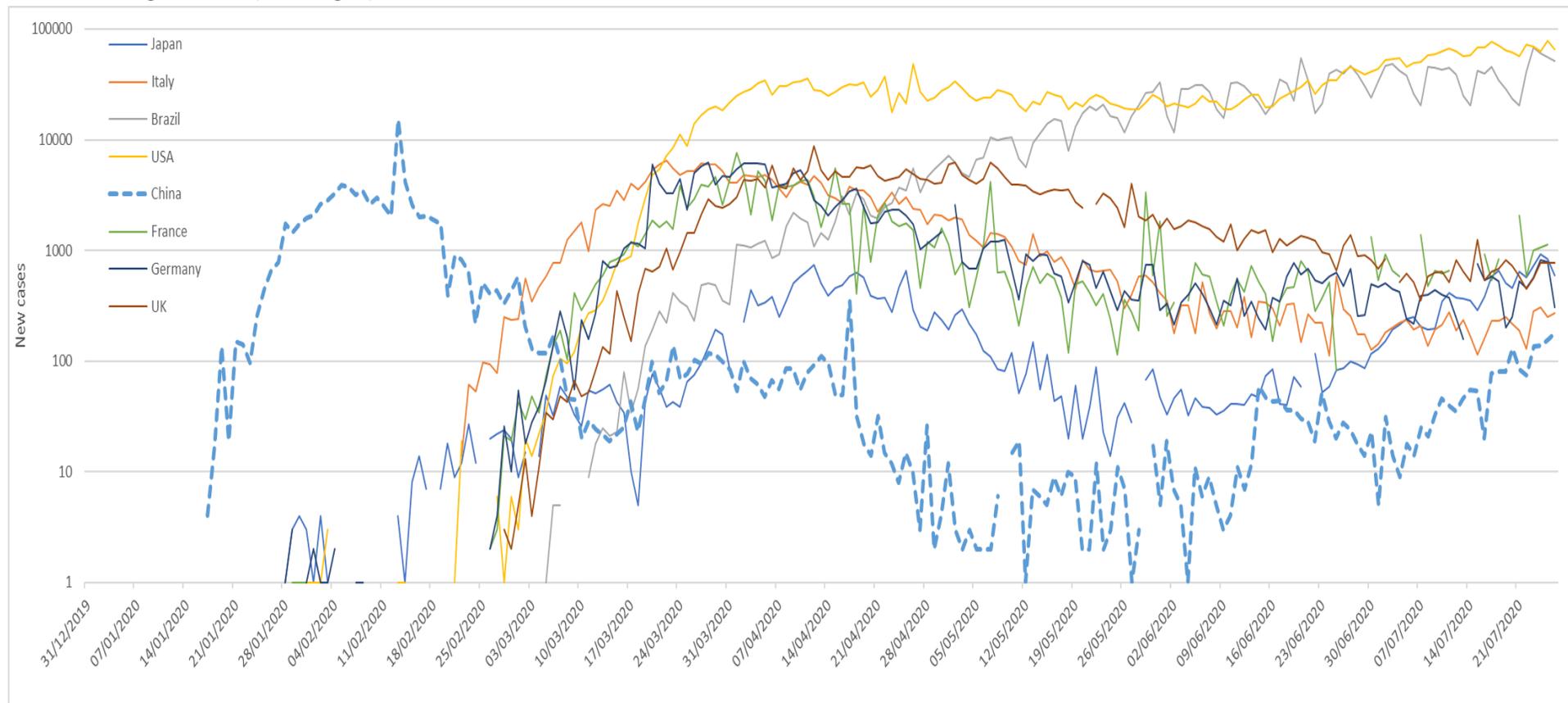

Fonte: elaborazioni su dati https://ourworldindata.org/coronavirus-source-data.

**2. Distribuzione prima cifra – Brasile**

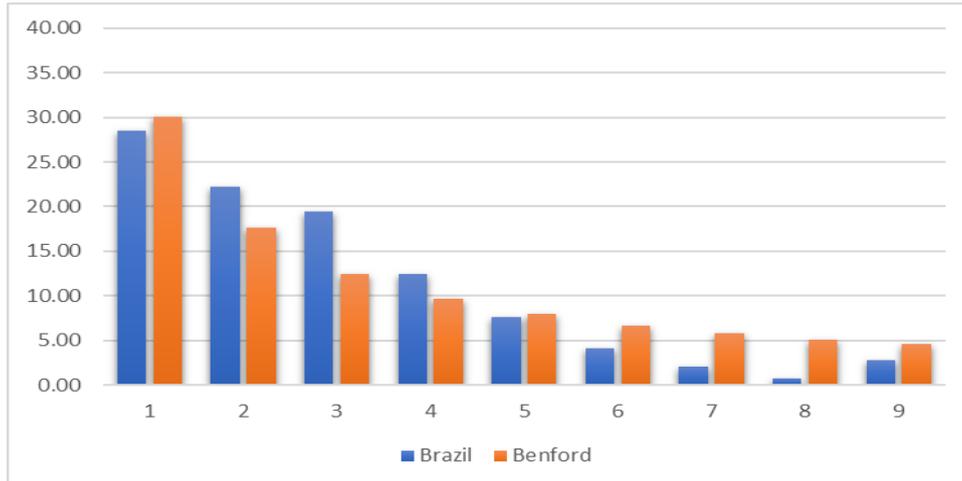

Fonte: elaborazioni su dati https://ourworldindata.org/coronavirus-source-data.

**3. Distribuzione prima cifra - Cina**

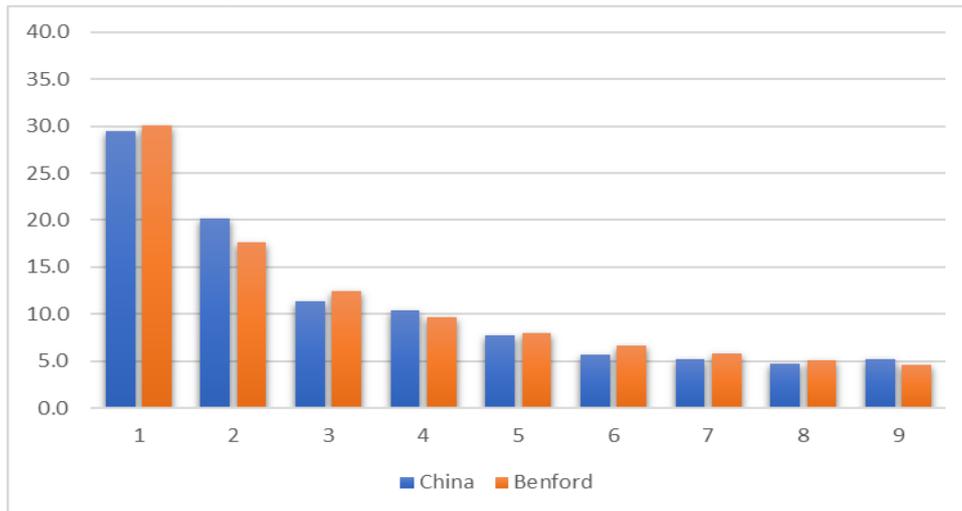

Fonte: elaborazioni su dati https://ourworldindata.org/coronavirus-source-data.

**4. Distribuzione prima cifra - Francia**

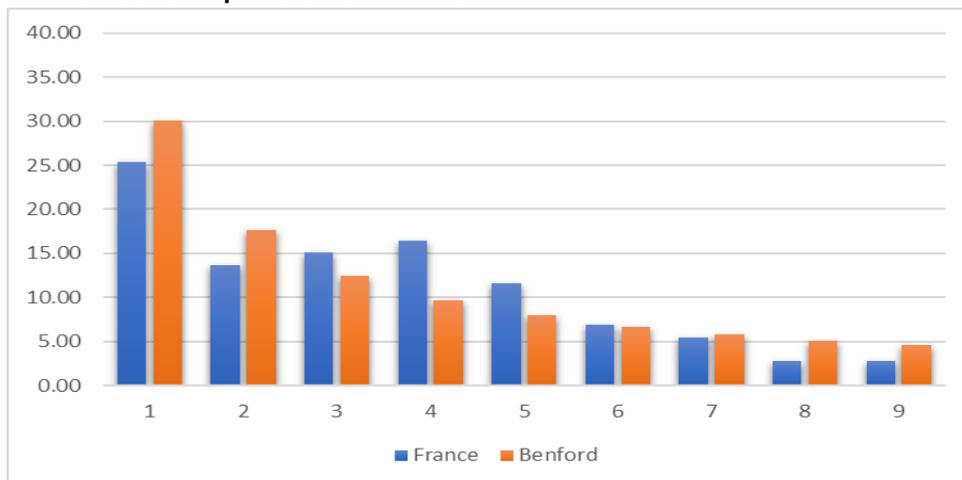

Fonte: elaborazioni su dati https://ourworldindata.org/coronavirus-source-data.

### 5. Distribuzione prima cifra - Germania

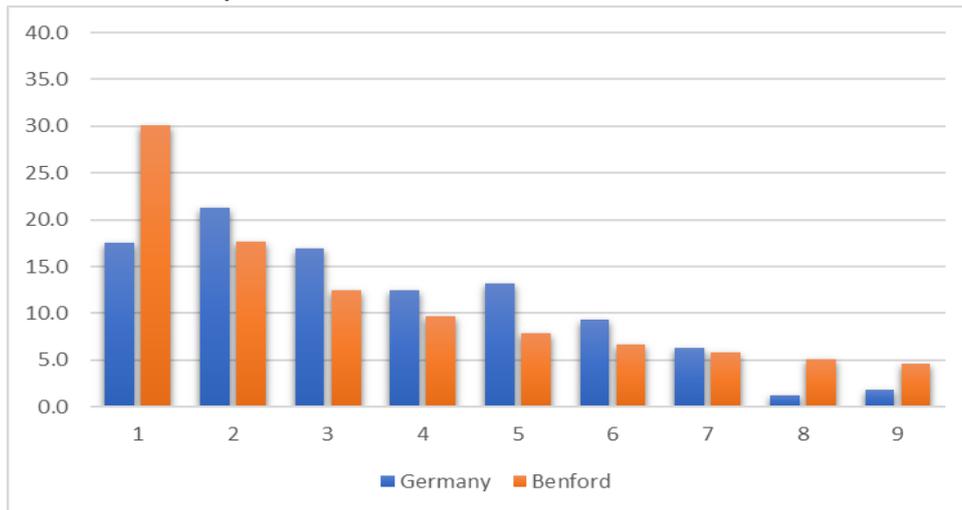

Fonte: elaborazioni su dati https://ourworldindata.org/coronavirus-source-data.

### 6. Distribuzione prima cifra - Giappone

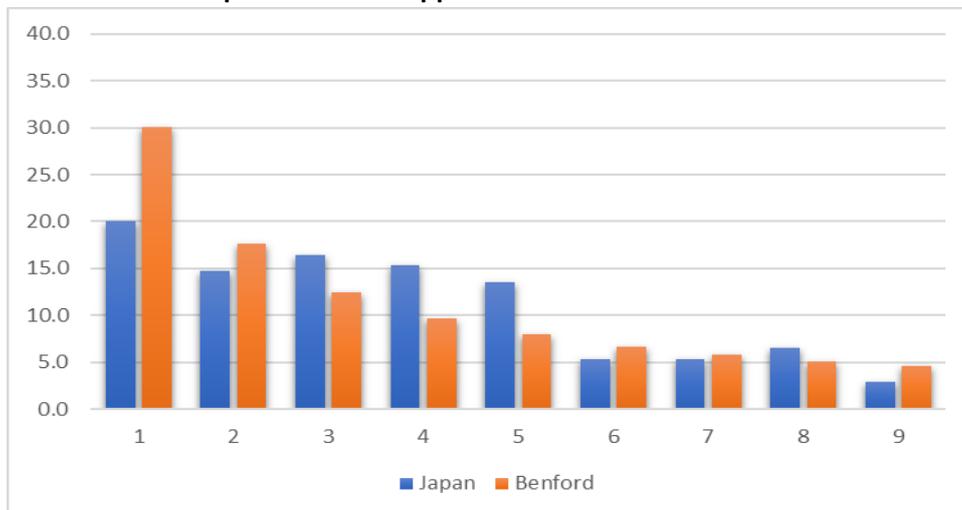

Fonte: elaborazioni su dati https://ourworldindata.org/coronavirus-source-data.

### 7. Distribuzione prima cifra - Italia

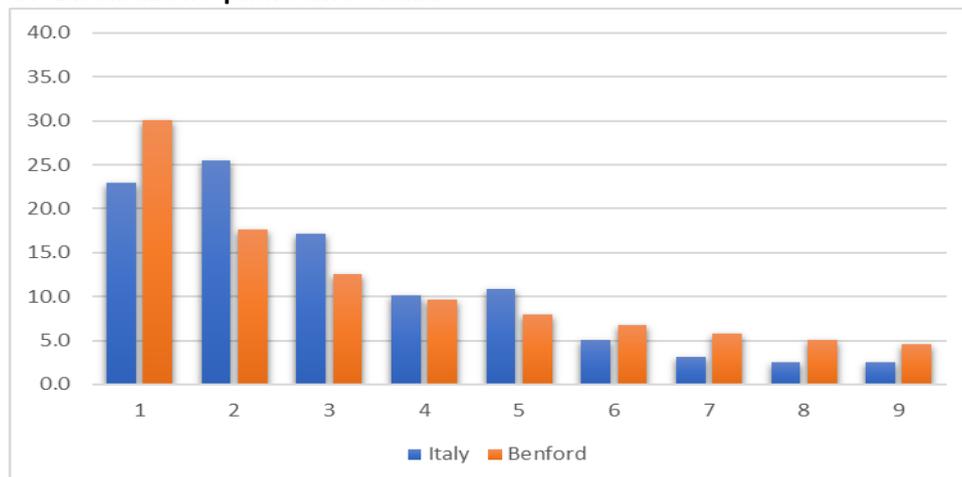

Fonte: elaborazioni su dati https://ourworldindata.org/coronavirus-source-data.

### 8. Distribuzione prima cifra – Regno Unito

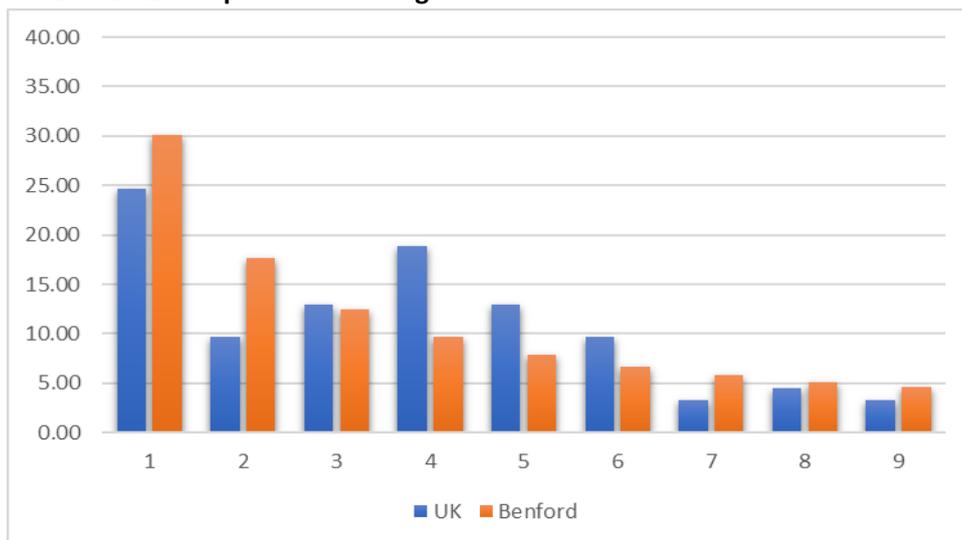

Fonte: elaborazioni su dati https://ourworldindata.org/coronavirus-source-data.

### 9. Distribuzione prima cifra - Stati Uniti

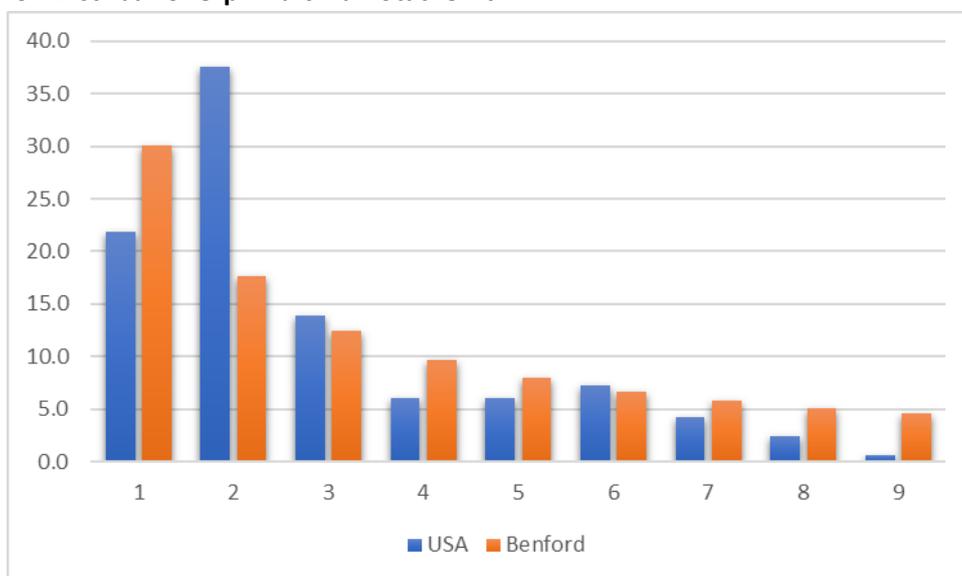

Fonte: elaborazioni su dati https://ourworldindata.org/coronavirus-source-data.

### 10. Test chi-quadro: confronto distribuzioni prime cifre Paesi vs distribuzione discreta uniforme (sx) e distribuzione Benford (dx)

|  | Nr. oss. | p-value | Sign. 95% | Sign. 99% |
|---|---|---|---|---|
| **Brasile** | 144 | 0.0000 |  |  |
| **Cina** | 193 | 0.0000 |  |  |
| **Francia** | 146 | 0.0000 |  |  |
| **Germania** | 160 | 0.0000 |  |  |
| **Giappone** | 170 | 0.0000 |  |  |
| **Italia** | 157 | 0.0000 |  |  |
| **Regno Unito** | 154 | 0.0000 |  |  |
| **Stati Uniti** | 165 | 0.0000 |  |  |

|  | Nr. oss. | p-value | Sign. 95% | Sign. 99% |
|---|---|---|---|---|
| **Brasile** | 144 | 0.0105 |  | * |
| **Cina** | 193 | 0.9892 | * | * |
| **Francia** | 146 | 0.0540 | * | * |
| **Germania** | 160 | 0.0005 |  |  |
| **Giappone** | 170 | 0.0031 |  |  |
| **Italia** | 157 | 0.0172 |  | * |
| **Regno Unito** | 154 | 0.0002 |  |  |
| **Stati Uniti** | 165 | 0.0000 |  |  |

Fonte: elaborazioni su dati https://ourworldindata.org/coronavirus-source-data.

**Osservazioni conclusive**

L'analisi qui condotta dimostra come questi dati sembrino rispettare la distribuzione delle prime cifre della legge di Benford. In particolare, per alcuni dei Paesi considerati (Brasile, Cina, Francia e Italia) vi sarebbe anche una evidenza statistica a rafforzare tale convinzione.

Quanto qui analizzato potrebbe essere utilizzato per capire l'andamento dei contagi. Lungi dal suggerire un modello predittivo, fuori dallo scopo solo descrittivo e analitico di questo lavoro, si può immaginare che i Paesi che sono ancora lontani dall'avere una distribuzione più aderente a quella di Benford siano più lontani dalla conclusione della pandemia. Per esempio, ciò potrebbe riferirsi soprattutto a casi come quelli di Stati Uniti e Regno Unito, sebbene un Paese in piena emergenza come il Brasile sembri già aderire alla distribuzione di Benford. Per verificare tale opportunità, si ritiene utile riverificare tra qualche mese la situazione, in modo da verificare se eventuali Paesi che avranno definitivamente superato l'epidemia avranno una distribuzione simile a quella immaginata.

La conclusione più evidente è, invece, che anche questa volta sembrerebbe essere confermato il contenuto di questa affascinante legge empirica.